\documentclass[onecolumn,nobibnotes,nofootinbib]{revtex4}
\usepackage{amsmath,amssymb}
\usepackage{graphicx,subfigure,epsfig}

\newcommand{\be}{\begin{equation}}
\newcommand{\ee}{\end{equation}}
\newcommand{\bea}{\begin{eqnarray}}
\newcommand{\eea}{\end{eqnarray}}
\newcommand{\benn}{\begin{displaymath}}
\newcommand{\eenn}{\end{displaymath}}
\newcommand{\beann}{\begin{eqnarray*}}
\newcommand{\eeann}{\end{eqnarray*}}
\newcommand{\oone}{
\begin{picture}(10,8)
\put(5,5){\circle{8}}
\put(2.9,2.5){{\scriptsize 1}}
\end{picture}
}

\newcommand{\beq}{\begin{equation}}
\newcommand{\eeq}{\end{equation}}
\newcommand{\nn}{\nonumber\\}
\newcommand{\dif}{{\rm d}}

\newcommand{\bx}{x_{\bot}}
\newcommand{\by}{y_{\bot}}
\newcommand{\bu}{z'_{\bot}}

\newcommand{\bz}{z_{\bot}}

\newcommand{\abar}{\bar{\alpha}_s}

\newcommand{\Nc}{N_{\rm c}}

\newcommand{\Nf}{N_{\rm f}}

\newcommand{\minus}{\!-\!}

\begin{document}

\title{Solution to the evolution equation at NLL for high parton density QCD}
\author{Wenchang Xiang$^{a,b,c}$}
\email{wxiangphy@gmail.com}
\author{Shaohong Cai$^a$}
\email{aa.shcai@gzu.edu.cn}
\author{Daicui Zhou$^b$}
\email{dczhou@mail.ccnu.edu.cn}
\affiliation{a) Department of Physics, Guizhou University, Center for Interdisciplinary Studies in Physics and Related Areas, Guizhou University of Finance and Economics, Guiyang 550025, China\\
b) Institute of Particle Physics, Central China Normal University, Wuhan 430079, China\\
c) Department of Physics, Guizhou Normal University, Guiyang 550001, China}
%\pacs{}
\date{\today}

\begin{abstract}
We analytically solve the full next-to-leading logarithmic Balitsky-Kovchegov equation in the saturation regime, which includes corrections from quark and gluon loops, and large double transverse logarithms. The analytic result for the $S$-matrix in the saturation regime shows that the linear rapidity decrease with rapidity of the exponent in running coupling case is replaced by rapidity raised to power of 3/2 decreasing. The collinearly-improved Balitsky-Kovchegov equation are also analytically solved in the saturation regime. It shows that the double collinear logarithms do not contribute to the $S$-matrix and the solution is the same as the one obtained from the leading order Balitsky-Kovchegov equation. The numerical solutions to the leading order and full next-to-leading logarithmic Balitsky-Kovchegov equations are performed in order to test the analytic results derived in the saturation regime.
\end{abstract}

\maketitle

%-----------------------------------------------------------------------------

\section{Introduction}
\label{sec:intro}
At high energy dipole-hadron scattering perturbative QCD (pQCD) predicts a rapid growth of the gluon density with increasing rapidity (or energy), which leads to non-linear phenomena such as gluon saturation and multiple scattering. The non-linear evolution of amplitude between dipole and hadron scattering is well established by the Balitsky-JIMWLK\footnote{The JIMWLK is the abbreviation of Jalilian-Marian, Iancu, McLerran, Weigert, Leonidov, Kovner.} hierarchy\cite{Ba,JIMWLK1,JIMWLK2,JIMWLK3,JIMWLK4} and its mean field approximation known as the Balitsky-Kovchegov (BK) equation\cite{Ba,K} at leading logarithmic accuracy. The BK equation resums leading logarithms $\sim \alpha_s\ln(1/x)$ to all orders, where $x$ is the Bjorken variable. The phenomenological studies of the experimental data from structure functions in deep inelastic scattering at HERA\cite{JS,JM}, and particle production in heavy ion collisions at LHC\cite{LM3,CX} have been found that the BK equation at leading logarithmic accuracy in pQCD is insufficient to describe these high energy scattering processes.

Over the past decade, our understanding of the BK equation beyond leading logarithmic accuracy has been improved due to including the running coupling corrections\cite{Bnlo, KW}. A running coupling BK (rcBK) equation is obtained, which resums all order corrections associated with the coupling and is an improved version of the leading order (LO) BK equation with the fixed coupling $\alpha_s$ in BK kernel replaced by a ¡°triumvirate¡± of the running couplings. The numerical studies on the solutions of the rcBK have been found that these corrections are numerically large and these corrections significantly slow down the evolution speed of the dipole amplitude with increasing rapidity. However, the running coupling is not the only large corrections to the LO BK equation. In addition to the running coupling corrections (in the language of Feynman diagram, contribution from quark loops), the authors in Ref.\cite{BC} have been found that the gluon loops also take a large contribution to the rapidity evolution of the dipole amplitude. Via including contributions from quark and gluon loops, Balitsky and Chirilli got a full next-to-leading (NLO) order BK equation. It is not hard to find that the double transverse logarithmic term in the full NLO BK equation is large when the dipole size $r$ is small, which leads to the equation enhanced by double (or collinear) logarithms, see Eq.(\ref{nlobk}). However, the full NLO BK equation suffers from a severe lack of stable problem. The solution of the full NLO BK equation
strongly depends on the details of the initial condition\cite{LM}. The solution can be negative with increasing rapidity, which is not a physically meaningful evolution. The origin of this instability can be traced back to the a large double transverse logarithmic NLO correction
($\sim \ln\frac{(x_{\bot}-z_{\bot})^2}{(x_{\bot}-y_{\bot})^2}\ln\frac{(y_{\bot}-z_{\bot})^2}{(x_{\bot}-y_{\bot})^2}$) in the full NLO BK equation\cite{LM}. To solve this instability problem, a resummation of the double transverse logarithms has to be performed.

Recently, the radiative corrections enhanced by double collinear logarithms are resummed by the authors in Ref.\cite{IM}. They resummed these corrections to all orders by solving a non-local evolution equation, which can be reformulated to a local equation in rapidity with modified kernel and initial condition. The instability behavior is solved once the double transverse logarithms are resummed to all orders. We denote the collinearly-improved evolution equation as the double logarithmic approximation (DLA) BK equation in this study. One can extend the DLA evolution equation to full next-to-leading logarithmic (NLL) accuracy by adding the NLO BK corrections from \cite{BC}. The numerical solutions to the full NLO, DLA and full NLL BK equations are studied in Refs.\cite{LM,IM,LM2}. The numerical results show the essential role of the double transverse logarithmic resummation in stabilizing. In addition, the dipole amplitude significantly is slowed down by the double logarithmic resummation. Although the numerical results can be directly applied to phenomenologies, like deep inelastic scattering\cite{IM2,JL}, single particle production and two-particle correlations in heavy ion collisions\cite{RA,LM3,SX}, they would be very cumbersome to use in practice due to the intricate. If the analytic solutions to these equations are available, then one can establish an elegant analytic dipole amplitude, like IIM amplitude\cite{IIM} in LO BK, which would be more convenient to use in phenomenology than the numerical one.

In our previous publication\cite{WX}, we analytically solved the rcBK equation in the saturation region. We found that the running coupling corrections modify the $S$-matrix a lot as compared to the fixed coupling case. The exponent in $S$-matrix is linear decrease with rapidity in the case of running coupling while the exponent in $S$-matrix decreases quadratically with rapidity in LO BK case, which indicates that the running coupling slows down the rapidity evolution of the scattering amplitude. In this work, we analytically solve the full NLO and NLL BK equations in the saturation region and obtain their analytic results for the $S$-matrix at high energies. We find that the DLA and LO BK equations have the same analytic solutions in the saturation region. Moreover, the full NLO and NLL BK equations also have the same solutions in the saturation regime. Interestingly, the analytic solution of the full NLL BK equation shows that the linear decrease with rapidity of the exponent in the solution to the rcBK equation is replaced by rapidity raised to power of 3/2, which indicates the gluon loop contributions partially compensate the reduction from the quark loops in the saturation region. To test these outcomes, we numerically solve the above equations with focusing on the saturation region. The numerical results support the analytic findings.

%------------------------------------------------------------------------------

\section{The LO, running coupling BK equations and their solutions}
\label{sec:BK_eq}
To motivate the higher order corrections and for comparison with more refined results which shall be obtained later, in this section we recall the low level (LO and rc) BK equations which describe the rapidity $Y=\ln(1/x)$ evolution of the $S$-matrix, $S(x_{\bot},y_{\bot},Y)$ = 1 -  $N(x_{\bot},y_{\bot},Y)$, of a $q\bar{q}$ dipole scattering off a target which may be another dipole, a hadron or a nucleus. The scattering amplitude is small in the dilute target region, while at the dense target regime it approaches the unitarity limit ($N \rightarrow 1$), which indicates the $S$-matrix approaching zero at the saturation region. For simplicity we analytically solve the BK equations in terms of the $S$-matrix instead of the scattering amplitude $N$ in the following studies.
%-----------------------------------------------------------
\subsection{The LO BK equation and its analytic solution }
\label{Sec_Kovchegov_equation}
%------------------------------------------------------------
The BK equation describes the rapidity evolution of the $S$-matrix of a quark-antiquark dipole (or a projectile) with a quark leg at transverse coordinate $x_{\bot}$ and an antiquark at transverse coordinate $y_{\bot}$ scattering off a hadronic target. To see how the $S$-matrix evolving with rapidity $Y$, we take $Y$ boosting a small increment $dY$. If we put the evolution at dipole framework and let the target fixed, then the dipole has a probability $dP$ to emit a gluon due to the increment of $dY$\cite{M},
\be
dP = \frac{\bar{\alpha}_s}{2 \pi}
     \frac{(x_{\bot}-y_{\bot})^2}{(x_{\bot}-z_{\bot})^2(z_{\bot}-y_{\bot})^2}d^2z_{\bot} dY,
\ee
where $z_{\bot}$ is transverse coordinate of the emitted gluon, and $\bar{\alpha}_s = \alpha_sN_c/\pi$.
For convenient later on calculations, we denote ${\bf r}=x_{\bot}-y_{\bot}$, ${\bf
r}_1=x_{\bot}-z_{\bot}$ and ${\bf r}_2=z_{\bot}-y_{\bot}$ as the
sizes of parent and of the new daughter dipoles produced by the
evolution, respectively. In the large $N_c$ limit, the quark-antiquark-gluon state
can be viewed as two daughter dipoles -- one of the dipoles consists of the initial quark and the antiquark part of the gluon
while the other dipole is constituted by the quark part of the gluon and
the initial antiquark. In the dipole framework, the increment of $S$-matrix, $dS$, due to the boosting of $dY$ can be written by
multiplying the probability $dP$ with the $S$-matrix\cite{M}
\be
\frac{\partial}{\partial Y} S(r, Y) =
           \frac{\bar{\alpha}_s}{2 \pi} \int d^2r_1
           \frac{r^2}{r_1^2r_2^2} \left [
           S^{(2)}(r_1, r_2,Y) -
           S(r, Y) \right ],
\label{eq_kov_1}
\ee
where $S^{(2)}(r_1, r_2, Y)$ expresses two daughter dipoles simultaneous
scattering off the target. The last term in~(\ref{eq_kov_1}) describes the scattering of a single dipole on the target.
Eq.(\ref{eq_kov_1}) is an integro-differential equation and gives the scattering amplitude at all rapidities $Y>0$. A few more words about the interpretation of the Eq.(\ref{eq_kov_1}) are in order. The energy evolution follows from the gluon emission becoming possible when the dipole is boosted to higher rapidity. Integration over the rapidity interval is corresponding to multiple gluon emissions and thus we have large number of dipoles in the projectile wave function. Eq.(\ref{eq_kov_1}) is an infinite hierarchy of coupled evolution equations due to lower number of dipoles scattering always coupling to higher number of dipoles, for instance two dipoles scattering coupling to three dipoles. Therefore, Eq.(\ref{eq_kov_1}) is almost impossible for direct applications to phenomenology, since $S^{(2)}$ is not known. To get a closed equation, we need to employ the large $N_c$ limit, the $S^{(2)}$ can be simplified to
\be
S^{(2)}(r_1, r_2, Y) =
  S(r_1, Y) S(r_2, Y),
\label{fac}
\ee
then we obtain the BK equation\cite{Ba,K}
\be
\frac{\partial}{\partial Y} S(r, Y) =
           \frac{\bar{\alpha}_s}{2 \pi} \int d^2r_1
           \frac{r^2}{r_1^2r_2^2} \left [
           S(r_1, Y)S(r_2, Y) -
           S(r, Y) \right ].
\label{loBKeq}
\ee

Now let's analytically solve the BK equation in the saturation region where the target density is large and the scattering is strong with $N\sim 1$, thus $S\sim 0$. The non-linear term in (\ref{loBKeq}) is negligible. In the logarithmic regime of integration, the Eq.(\ref{loBKeq}) becomes
\be
\frac{\partial}{\partial Y} S(r, Y) =
           - 2\frac{\bar{\alpha}_s}{2 \pi}\pi\int^{r^2}_{1/Q_s^2}
           dr_1^2
           \frac{1}{r_1^2}
           S(r, Y)\ .
\label{Eq_Kovchegov_S}
\ee
Note that we work in the saturation region in which the dipole size is much larger than characteristic size $1/Q_s$, the $Q_s$ is saturation momentum which controls the separation between dilute and dense regimes. Thus, the lower and upper bounds of integration in (\ref{Eq_Kovchegov_S}) are $1/Q_s^2$ and $r^2$, respectively. The factor 2 in (\ref{Eq_Kovchegov_S}) comes from the symmetry of the two regions dominating the integral, either from $1/Q_s\ll \mid\mathbf{r_1}\mid\ll \mid\mathbf{r}\mid,~\mid\mathbf{r_2}\mid\sim \mid\mathbf{r}\mid$ or $1/Q_s\ll \mid\mathbf{r_2}\mid\ll \mid\mathbf{r}\mid,~\mid\mathbf{r_1}\mid\sim \mid\mathbf{r}\mid$, see Figure 1. Via performing the integrations over transverse coordinate and rapidity, we obtain the analytic solution of the BK equation\cite{HM,LT}
\be
S(r, Y)=\exp\left[-\frac{c}{2}\bar{\alpha}_s^2(Y-Y_0)^2\right]S(r, Y_0)
\label{Sol_Kovchegov},
\ee
with $Q_s^2(Y)=\exp\left[c\bar{\alpha}_s(Y-Y_0)\right]Q_s^2(Y_0)$ and $Q_s^2(Y_0)r^2=1$. One can see that the exponent has a quadratic rapidity decrease.  

The analytic solution to the BK equation in the saturation region has been found by
Levin and Tuchin~\cite{LT}. The reason why we have gone through such a
detailed ``derivation'' of (\ref{Sol_Kovchegov}) since one of the main purposes of this study is to
show how the leading order $S$-matrix is modified by the higher order contributions like running coupling, full NLO and large double transverse logarithmic corrections.
\begin{figure}[h!]
\setlength{\unitlength}{1.5cm}
\begin{center}
\epsfig{file=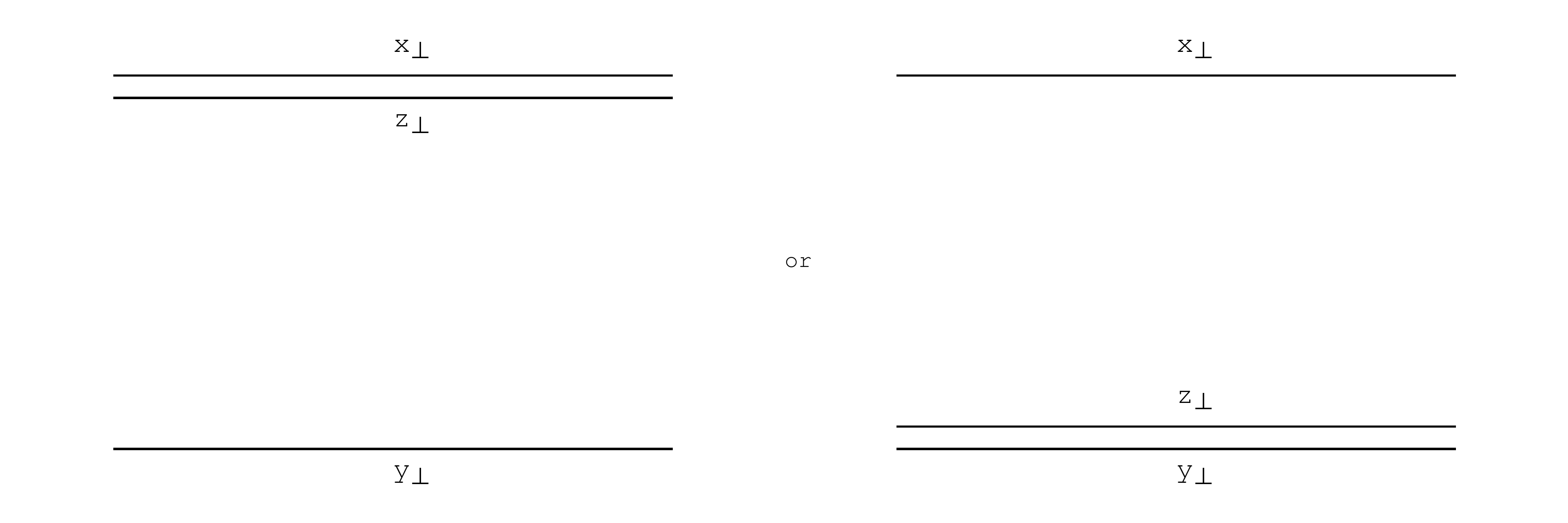, width=15.5cm,height=5cm}
\end{center}
\caption{The coordinates of dipoles and saturation region.}
\label{nlo}
\end{figure}
%

%---------------------------------------------------------------------------------------------

%--------------------------------------------------------------
\subsection{The running coupling BK equation and its analytic solution}
\label{rcBKeqsec}

Beyond the leading logarithmic approximation, the authors in Refs.\cite{KW,Bnlo} calculated the higher order perturbative corrections by including fermion (quark) bubble diagrams which bring in a factor of $\alpha_sN_f$ and modify the BK evolution kernel. Once including the higher order corrections there are two impacts on
the Feynman diagrams which describe the dipole structure generated under evolution.
First, the propagator of the emitted gluon, which comes from the emission of the original parent dipole when the parent dipole boosts to higher rapidities, is now dressed with quark loops in contrast to leading order or fixed coupling case, which will modify the emission probability of gluon (or BK kernel), but keeping the leading order interaction terms untouched. Second, the quark-antiquark pair is added to the evolved wave function, which not only modifies the BK kernel but also changes the interaction structure of the evolution equation. In this case the evolution of the $S$-matrix is proportional to the product of two $S$-matrixes of the newly created dipoles. Concerning two above aspects and resumming $\alpha_s N_f$ to all orders, one can get the running coupling BK equation\cite{JA}
\bea
\frac{\partial S(x_{\bot}-y_{\bot},Y)}{\partial Y}= \int\,d^2 z_{\bot}
  \,\tilde{K}(x_{\bot},y_{\bot}, z_{\bot})
  \left[S(x_{\bot}-z_{\bot}Y)\,S(z_{\bot}-y_{\bot},Y)-S(x_{\bot}-y_{\bot},Y)\right]\nonumber \\
   - \alpha_{\mu}^2 \, \int d^2 z_{{\bot}1} \, d^2 z_{{\bot}2} \,
  K_{\oone} (x_{\bot}, y_{\bot} ; z_{{\bot}1}, z_{{\bot}2}) \, \left[ S (x_{\bot}-
   w_{\bot}, Y) \, S (w_{\bot}-y_{\bot}, Y)\right. \nonumber \\
   \left. - S (x_{\bot}-z_{{\bot}1}, Y) \, S (z_{{\bot}2}-y_{\bot}, Y)\right],
\label{BK_RC1}
\eea
where $w_{\bot}$ is the point of subtraction in the coordinate space, and $z_{\bot}$ is the transverse coordinate of the emitted gluon. In the large $N_c$ limit, the emitted gluon can be viewed as a quark-antiquark pair with a quark leg at transverse coordinate $z_{{\bot}1}$ and an antiquark leg at transverse coordinate $z_{{\bot}2}$.  The first line on the r.h.s of (\ref{BK_RC1}) refers to as the 'running coupling' contribution and
resums all power of $\alpha_s N_f$ corrections to all orders. It has the same structure as the leading order BK equation
but with modified kernel due to the running coupling corrections. The modified kernel $\tilde{K}(x_{\bot},y_{\bot}, z_{\bot})$
has two kinds of expressions\cite{Bnlo,KW}, which depends on the scheme choice (see~\cite{JA} for more discussions on the
scheme choice). We adopt the choice derived by Balitsky in Ref.\cite{Bnlo}. The kernel of the running coupling BK can be
written as\cite{Bnlo}
\be
  K^{rc}({\bf r},{\bf r}_1,{\bf r}_2)=\frac{\bar{\alpha}_s(r^2)}{2\pi}
  \left[\frac{r^2}{r_1^2\,r_2^2}+
    \frac{1}{r_1^2}\left(\frac{\alpha_s(r_1^2)}{\alpha_s(r_2^2)}-1\right)+
    \frac{1}{r_2^2}\left(\frac{\alpha_s(r_2^2)}{\alpha_s(r_1^2)}-1\right)
  \right]\,.
\label{kbal}
\ee
 We would like to note that the kernel, $K_{\oone} (x_{\bot}, y_{\bot} ; z_{{\bot}1}, z_{{\bot}2})$, in the second line on the r.h.s of~(\ref{BK_RC1}) is not discussed in this work (see \cite{JA} for relevant discussion), since it corresponds to quadratic terms of $S$-matrix which are negligible in current study.

In the saturation regime in which the interaction between dipole and hadron
is very strong, $S(x_{\bot}-y_{\bot},Y)\rightarrow 0$, and
unitarity corrections become important, the quadratic terms
in~(\ref{BK_RC1}) can be neglected in which case one needs only
keep the second term in the first line on the r.h.s of~(\ref{BK_RC1}). The evolution
equation including running coupling is given by
\be
\frac{\partial S(r, Y)}{\partial Y}=-\int\,d^2 r_1
  \,K^{rc}({\bf r},{\bf r}_1,{\bf r}_2)S(r, Y).
\label{SKW_sr}
\ee
The expert reader will recognize that Eq.(\ref{SKW_sr}) has the same form as the LO BK Eq.(\ref{Eq_Kovchegov_S})
with only the kernel $\tilde{K}({\bf r},{\bf r}_1,{\bf r}_2)$ modified by running coupling corrections. We use similar approach as the solution to Eq.(\ref{Eq_Kovchegov_S}) to analytically solve the rcBK equation.

In the saturation region, the main contribution to the integration on the r.h.s of~(\ref{SKW_sr}) comes from either
$1/Q_s\ll r_1\ll r, r_2\sim r$ or $1/Q_s\ll r_2\ll r, r_1\sim r$ region, see Figure 1.
We choose the first regime, $1/Q_s\ll r_1\ll r, r_2\sim r$. Under this approximation, the kernel becomes
\bea
K^{rc}({\bf r},{\bf r}_1,{\bf r}_2)&=& \frac{\abar(r^2)}{2\pi}\left[\frac{1}{r_1^2}\frac{\alpha_s(r_1^2)}
{\alpha_s(r^2)}+\frac{1}{r^2}\left(\frac{\alpha_s(r^2)}{\alpha_s(r_1^2)}-1\right)\right]\nonumber\\
&\simeq&\frac{\abar(r_1^2)}{2\pi r_1^2},
\label{balk}
\eea
with the running coupling at one loop accuracy
\be
\abar(r_1^2)=\frac{N_c}{\pi}\frac{\mu}{1 + \mu_1\ln\left(\frac{1}{r_1^2\Lambda^2}\right)}.
\ee

Substituting the simplified kernel~(\ref{balk}) into~(\ref{SKW_sr}),
we can get rcBK equation in the saturation region as:
\be
\frac{\partial S(r,Y)}{\partial Y}=-2\frac{1}{2\pi}\int_{1/Q_s^2}^{r^2}\,d^2 r_1
  \,\frac{\abar(r_1^2)}{r_1^2}S(r,Y),
\label{SKW_f}
\ee
whose solution is\cite{WX}
\bea
S(r,Y) &=& \exp\left[-\frac{N_c\mu}{c\pi\mu_1}\left(\ln^2\left(\frac{Q_s^2(Y)}{\Lambda^2}\right)\ln\left(\frac{1+\mu_1\ln\frac{Q_s^2}
{\Lambda^2}}{1+\mu_1\ln\frac{1}{r^2\Lambda^2}}-\frac{1}{2}\right)+\frac{1}{\mu_1}\ln\left(\frac{Q_s^2(Y)}{\Lambda^2}\right)
\right.\right.\nn
 &&\hspace*{0.8cm}-\left.\left.\frac{1}{\mu_1^2}\ln\left(1+\mu_1\ln\frac{Q_s^2}{\Lambda^2}\right)\right)\right]S(r_0,Y)
\label{solution_running}
\eea
with
\be
\ln(Q_s^2(Y)/\Lambda^2)=\sqrt{c(Y-Y_0)}+\textsl{O}(Y^{1/6}).
\ee
Comparing to the $S$-matrix in the LO case, the $S$-matrix in (\ref{solution_running}) is modified by the running coupling corrections. The quadratic rapidity decrease in exponent at LO BK (see Eq.~(\ref{Sol_Kovchegov})) is replaced by linear decrease with rapidity in (\ref{solution_running}), which indicates that the scattering amplitude is slowed down by the running coupling corrections. This result is in agreement with theoretic expectations\cite{Bnlo}.

We wish to note that although the running coupling kernel depends on the scheme choice, if one chooses the Kovchegov-Weigert scheme instead of Balitsky scheme and will find that the analytic solution to the rcBK is independent of the scheme choice. In other words, the running coupling Balitsky and Kovchegov-Weigert evolution equations give the same analytic solution in the saturation region.
These two running coupling evolution equations are equivalent to each other in the saturation region.

%--------------------------------------------------------------------------

\section{The full NLO, DLA, full NLL BK equations and their solutions}
In the last section, we discuss the running coupling corrections to the LO BK equation. However, the running coupling (or bubble chain of quark loops) is not the only higher order perturbative corrections to the LO BK equation. The gluon loops also bring a large contributions to the LO BK equation. The full NLO BK equation can be derived by including contributions from the quark and gluon loops as well as from the tree gluon diagrams with quadratic and cubic nonlinearities\cite{BC}.

\subsection{Full NLO BK equation and its analytic solution}
We analytically solve the full NLO BK evolution equation derived by Balitsky and Chirilli in Ref.\cite{BC}, which can be written as:
 \begin{align}
 \label{nlobk}
 \hspace*{-0.7cm}
 \frac{\partial S(\bx-\by, Y)}{\partial Y} = \,&
 \frac{\abar}{2 \pi}
 \int \dif^2 \bz \,
 \frac{(\bx\minus\by)^2}{(\bx \minus\bz)^2 (\by \minus \bz)^2}\,
 \bigg\{ 1 + \frac{\abar}{4}
 \bigg[b\, \ln (\bx \minus \by)^2 \mu^2
 - b\,\frac{(\bx \minus\bz)^2 - (\by \minus\bz)^2}{(\bx \minus \by)^2}
 \ln \frac{(\bx \minus\bz)^2}{(\by \minus\bz)^2}
 \nn
 &
 +\frac{67}{9} - \frac{\pi^2}{3} - \frac{10 \Nf}{9 \Nc}-
 2\ln \frac{(\bx \minus\bz)^2}{(\bx \minus\by)^2} \ln \frac{(\by \minus\bz)^2}{(\bx \minus\by)^2}\bigg]
 \bigg\} \left[S(\bx-\bz, Y) S(\bz-\by, Y) - S(\bx-\by, Y) \right]
 \nn
  +\, & \frac{\abar^2}{8\pi^2}
 \int \frac{\dif^2 \bz\,\dif^2 \bu}{(\bu \minus \bz)^4}
 \bigg\{-2
 + \left[\frac{ (\bx \minus \bz)^2 (\by \minus \bu)^2 + (\bx \minus\bu)^2 (\by \minus\bz)^2
- 4 (\bx \minus \by)^2 (\bu \minus \bz)^2}{(\bx \minus \bz)^2 (\by \minus \bu)^2 - (\bx \minus \bu)^2 (\by  \minus \bz)^2}\right.
 \nn
 & \hspace*{0.1cm}
\left. + \frac{(\bx \minus \by)^2 (\bu \minus \bz)^2}{(\bx \minus \bz)^2 (\by  \minus \bu)^2}
 + \frac{(\bx \minus \by)^4 (\bu \minus \bz)^4}{(\bx \minus \bz)^2 (\by  \minus \bu)^2((\bx \minus \bz)^2 (\by \minus \bu)^2 - (\bx \minus \bu)^2 (\by  \minus \bz)^2)}\right]
 \nn
 & \hspace*{0.1cm} \times
 \ln \frac{(\bx \minus \bz)^2 (\by \minus \bu)^2}{(\bx \minus \bu)^2 (\by  \minus \bz)^2}\bigg\}
 \left[S(\bx-\bu, Y) S(\bu-\bz, Y) S(\bz-\by, Y) - S(\bx-\bu, Y) S(\bu-\by, Y)\right]
 \nn
 +\, & \frac{\abar^2}{8\pi^2}\,
 \frac{\Nf}{\Nc}
 \int \frac{\dif^2 \bu \,\dif^2 \bz}{(\bu \minus \bz)^4}
 \bigg[2
 - \frac{(\bx \minus\bu)^2 (\by \minus\bz)^2 +
 (\bx \minus \bz)^2 (\by \minus \bu)^2
 - (\bx \minus \by)^2 (\bu \minus \bz)^2}{(\bx \minus \bz)^2 (\by \minus \bu)^2 - (\bx \minus \bu)^2 (\by  \minus \bz)^2}
  \nn
 & \hspace*{0.1cm} \times
 \ln \frac{(\bx \minus \bz)^2 (\by \minus \bu)^2}{(\bx \minus \bu)^2 (\by  \minus \bz)^2} \bigg]
 \left[S(\bx-\bz, Y) S(\bu-\by, Y)- S(\bx-\bu, Y) S(\bu-\by, Y) \right],
 \end{align}
where $\mu$ is the renormalization scale, $b=(11N_c - 2N_f)/3N_c$ is the first coefficient of the $\beta$ function and $N_f$ is the number of flavors. The full NLO BK equation, Eq.(\ref{nlobk}), shows several remarkable characteristics as compared to the LO BK equation.
\begin{itemize}
  \item The kernel in the first integration receives a correction of order $\mathcal{O}(\bar{\alpha}^2_s)$ compared to the LO BK kernel. Especially, the terms proportional to $b$ come from quark loop contributions (or running coupling corrections). The rest terms contain contributions from gluon loops, which include corrections enhanced by 'double logarithms'. The double logarithms become large in the small dipole size compared to the saturation scale $1/Q_s(Y)$, which can lead to negative value of scattering amplitude $N<0$\cite{LM}.
  \item The second and third integration terms are of order $\mathcal{O}(\bar{\alpha}^2_s)$. The double integrations over $\bz$ and $\bu$ refer to two additional partons (except for original quark and antiquark) with transverse coordinate $\bz$ and $\bu$ at the time of scattering.
  \item The second integration does not involve the number of quark flavor, $N_f$, which indicates that both two additional partons are gluons. This integration term describes the following sequence of evolutions: First the original quark-antiquark dipole ($\bx, \by$) radiates a gluon with transverse coordinate $\bu$, producing two daughter dipoles ($\bx, \bu$) and ($\bu, \by$); following the dipole ($\bu, \by$) emits another gluon with transverse coordinate $\bz$, thus creating the dipoles ($\bu, \bz$) and ($\bz, \by$). The cubic term $S(\bx-\bu, Y) S(\bu-\bz, Y) S(\bz-\by, Y)$ describes the two daughter gluons simultaneous interacting with the target, while the quadratic term $- S(\bx-\bu, Y) S(\bu-\by, Y)$ expresses the gluon at $\bz$ absent at the time of scattering.
  \item The third integration is proportional to the number of quark flavor, $N_f$, which indicates that the additional partons are a quark and an antiquark at the time of scattering. Otherwise, the evolutions are the same as the second integration case.
\end{itemize}

The running coupling terms in (\ref{nlobk}) are scale dependent, we use the scheme developed in \cite{Bnlo} to rewrite running coupling part and replace all terms proportional to $b$ by the Balitsky running coupling. The kernel in the first integration becomes:
\bea
\label{fnlokernel}
K^{fNLO}({\bf r},{\bf r}_1,{\bf r}_2) = \frac{\abar(r^2)}{2 \pi} \left\{\frac{r^2}{r_1^2r_2^2} + \frac{1}{r_1^2}\left[\frac{\alpha_s(r_1^2)}{\alpha_s(r_2^2)} - 1\right] + \frac{1}{r_2^2}\left[\frac{\alpha_s(r_2^2)}{\alpha_s(r_1^2)} - 1\right]\right.\nonumber\\
 \left.+ \hspace*{0.1cm} \frac{\bar{\alpha}_s(r^2)}{4}\frac{r^2}{r_1^2r_2^2}\left[\frac{67}{9} - \frac{\pi^2}{3} - \frac{10N_f}{9N_c} - 2\ln\frac{r_1^2}{r^2}\ln\frac{r_2^2}{r^2}\right]\right\}.
\eea

We analytically solve Eq.(\ref{nlobk}) in the saturation region where the scattering is strong, thus the scattering amplitude approaches unit ($N \sim 1$), in other words the $S$-matrix closes to zero at this regime. Therefore, the non-linear terms in (\ref{nlobk}) can be neglected in the saturation region. One finds that the full NLO BK equation reduces to
\be
\label{fnlobk}
 \frac{\partial S(r, Y)}{\partial Y} =
 - \int \dif^2 r_1 K^{fNLO}({\bf r},{\bf r}_1,{\bf r}_2) S(r, Y).
\ee
Note that in the saturation region the full NLO BK equation (\ref{fnlobk}) has similar structure as the rcBK equation (\ref{SKW_sr}) as well as LO BK equation (\ref{Eq_Kovchegov_S}), but with the kernel modified by quark and gluon loops.
To solve Eq.(\ref{fnlobk}), we can choose saturation region either $1/Q_s\ll r_1\ll r, r_2\sim r$ or $1/Q_s\ll r_2\ll r, r_1\sim r$, see Figure 1. In order to compare with running coupling case, we select the first region, $1/Q_s\ll r_1\ll r, r_2\sim r$. The $K^{fNLO}({\bf r},{\bf r}_1,{\bf r}_2)$ becomes
\bea
\label{sfnlo_kernel}
K^{fNLO}({\bf r},{\bf r}_1,{\bf r}_2) &=& \frac{\abar(r^2)}{2 \pi}\left[\frac{1}{r_1^2}\frac{\alpha_s(r_1^2)}{\alpha_s(r^2)} +
\frac{1}{r^2}\left(\frac{\alpha_s(r^2)}{\alpha_s(r_1^2)} - 1\right) + \frac{\bar{\alpha}_s(r^2)}{4}\frac{1}{r_1^2}\left(\frac{67}{9} - \frac{\pi^2}{3} - \frac{10N_f}{9N_c} - 2\ln\frac{r_1^2}{r^2}\ln\frac{r_2^2}{r^2}\right)\right]\nonumber\\
&\simeq& \frac{\abar(r_1^2)}{2\pi r_1^2} +
\frac{\bar{\alpha}^2_s(r^2)}{8\pi r_1^2}\left(\frac{67}{9} - \frac{\pi^2}{3} - \frac{10N_f}{9N_c}\right).
\eea
In the above equation, we neglect the second term in the first line on the r.h.s due to $r^2$ much larger than $r_1^2$. The double logarithmic term also can be neglected. One can see that the second term in the second line on the r.h.s of (\ref{sfnlo_kernel}) comes from the gluon loop correction.

Inserting the reduced kernel (\ref{sfnlo_kernel}) into (\ref{fnlobk}), the full NLO BK equation in the saturation region can be written as
\be
\label{fnlobk1}
 \frac{\partial S(r, Y)}{\partial Y} =
 - 2\frac{1}{2\pi}\int_{1/Q_s^2}^{r^2} \dif^2 r_1 \left[\frac{\abar(r_1^2)}{r_1^2} +
\frac{\bar{\alpha}^2_s(r^2)}{4}\frac{1}{r_1^2}\left(\frac{67}{9} - \frac{\pi^2}{3} - \frac{10N_f}{9N_c}\right)\right] S(r, Y),
\ee
and has the following solution
\bea
\label{sol_nll}
S(r, Y) &=& \exp\left[-\frac{N_c\mu}{2c\pi\mu_1}\left(\frac{2C_r}{3}\ln^3\frac{Q_s^2(Y)}
{\Lambda^2} + \ln^2\frac{Q_s^2(Y)}{\Lambda^2}\ln\frac{(r^2\Lambda^2)^{C_r} + (r^2\Lambda^2)^{C_r}\mu_1\ln\frac{Q_s^2(Y)}{\Lambda^2}}
{1 + \mu_1\ln\frac{1}{(r^2\Lambda^2)}}\right.\right.\nn
 &&\hspace*{0.8cm} + \left.\left.\frac{1}{\mu_1}\ln\frac{Q_s^2(Y)}{\Lambda^2} - \frac{1}{\mu_1^2}\ln\left(1+\mu_1\ln\frac{Q_s^2(Y)}{\Lambda^2}\right)\right)\right]
\eea
with $\ln(Q_s^2(Y)/\Lambda^2)=\sqrt{c(Y-Y_0)}+\textsl{O}(Y^{1/6})$ and $C_r = \alpha_s^2(r^2)N_c\mu_1(67/9 - \pi^2/3 - 10N_f/9N_c)/4\pi\mu$. Interestingly, the linear rapidity decrease in the exponent, (\ref{solution_running}), is now replaced by the rapidity raised to the power of 3/2, which indicates that gluon loop corrections compensate part of the decrease made by quark loops in the saturation region. In other words, the evolution speed of the scattering amplitude is slowed down in the full NLO BK compared to LO BK (quadratic rapidity decrease, see Eq.(\ref{Sol_Kovchegov})), but it is decreasing not as much as the rcBK case.

%------------------------------------------------------------

\subsection{Full NLL, DLA BK equations and their analytic solutions}
The numerical study of the full NLO BK equation found that its solution can turn to negative\cite{LM}, which is not meaningful in physics. This disaster can be traced back to the double logarithmic term which has to be resummed. By an explicit computation of the Feynman diagrams in light cone perturbation theory, the authors in Ref.\cite{IM} established an effective method to resum double logarithms to all orders and got a DLA BK equation which solves the problem of negative value of the scattering amplitude. The numerical solutions to the DLA BK equation show that the resummation of large double logarithmic corrections plays a significant role in stabilizing and slowing down the evolution. However, the DLA equation does not include contributions from quark and gluon loop corrections. It is possible to promote the DLA equation to full NLL accuracy by adding the NLO BK corrections derived by Balitsky and Chirilli in \cite{BC}. The kernel including double transverse logarithms (DTL) and full NLO corrections can be written as:
\bea
\label{knll}
K^{NLL}({\bf r},{\bf r}_1,{\bf r}_2)&=& \frac{\abar(r^2)}{2\pi}K^{DLA}\left[\frac{r^2}{r_1^2r_2^2} + \frac{1}{r_1^2}\left(\frac{\alpha_s(r_1^2)}{\alpha_s(r_2^2)} - 1\right)
+ \frac{1}{r_2^2}\left(\frac{\alpha_s(r_2^2)}{\alpha_s(r_1^2)} - 1\right)\right] \nn
 && + \frac{\abar^2(r^2)}{8\pi}\frac{r^2}{r_1^2r_2^2}\left[\frac{67}{9} - \frac{\pi^2}{3} - \frac{10N_f}{9N_c}\right],
\eea
with the DLA kernel\cite{IM}
\be
K^{DLA}(\rho) = \frac{J_1\left(2\sqrt{\abar\rho^2}\right)}{\sqrt(\abar\rho^2)} \simeq 1 - \frac{\abar\rho^2}{2} + \mathcal{O}(\abar^2),
\ee
where $J_1$ is the Bessel function, and $\rho = \sqrt{\ln r_1^2/r^2 \ln r_2^2/r^2}$. Note that the explicit double logarithmic term as the one in (\ref{fnlokernel}) is disappeared in (\ref{knll}), since it has already resummed into the kernel $K^{DLA}$.

Let's analyze the DLA kernel in the saturation region where $1/Q_s\ll r_1\ll r, r_2\sim r$ or $1/Q_s\ll r_2\ll r, r_1\sim r$, see Figure 1. To keep the consistency with previous two sections, we work in the region $1/Q_s\ll r_1\ll r, r_2\sim r$ at this section. The $\rho$ tends towards zero when $r_2 \sim r$, thus $K^{DLA} \simeq 1$, which corroborates the statement that the double logarithm only important in phase-space where the scattering is weak\cite{IM}. In the saturation region, the double logarithmic corrections can be neglected. The $K^{NLL}$  becomes
\be
\label{knlls}
K^{NLL}({\bf r},{\bf r}_1,{\bf r}_2) \simeq  \frac{\abar(r^2)}{2\pi r_1^2} + \frac{\abar^2(r^2)}{8\pi r_1^2}\left(\frac{67}{9} - \frac{\pi^2}{3} - \frac{10N_f}{9N_c}\right),
\ee
which is the same as the simplified full NLO kernel in (\ref{sfnlo_kernel}). The same kernel implies that the NLL BK equation in saturation region shares the exact the same solution as the full NLO BK equation. Its solution can be written as
\bea
S(r, Y) &=& \exp\left[-\frac{N_c\mu}{c\pi\mu_1}\left(\frac{2C_r}{3}\ln^3\frac{Q_s^2(Y)}
{\Lambda^2} + \ln^2\frac{Q_s^2(Y)}{\Lambda^2}\ln\frac{(r^2\Lambda^2)^{C_r} + (r^2\Lambda^2)^{C_r}\mu_1\ln\frac{Q_s^2(Y)}{\Lambda^2}}
{1 + \mu_1\ln\frac{1}{(r^2\Lambda^2)}}\right.\right.\nn
 &&\hspace*{0.8cm} + \left.\left.\frac{1}{\mu_1}\ln\frac{Q_s^2(Y)}{\Lambda^2} - \frac{1}{\mu_1^2}\ln\left(1+\mu_1\ln\frac{Q_s^2(Y)}{\Lambda^2}\right)\right)\right].
\eea

 Let's discuss the solution to the DLA BK equation derived in \cite{IM}
\be
\frac{\partial}{\partial Y} S(r, Y) =
           \frac{\bar{\alpha}_s}{2 \pi} \int d^2 r_1
           \frac{r^2}{r_1^2r_2^2} K^{DLA} \left [
           S(r_1, Y) S(r_2, Y) -
           S(r, Y) \right ].
\label{BKeq}
\ee
As we know the $K^{DLA} \sim 1$ in the saturation region, the solution of the DLA BK should be the same as the LO BK equation, see (\ref{Sol_Kovchegov}). We can see that the double transverse logarithmic correction is trivial in the saturation region and can be neglected, which is in agreement with the theoretical expectations\cite{IM}. In the next section, we numerically solve
the LO, DLA, rc and NLL BK equations to test the analytic results.

%-----------------------------------------------------------------------------
\section{Numerical Solution}
To test the outcomes what we get in the above sections for the analytic solutions to the LO, full NLO, and full NLL BK equations. We numerically solve these equations in this section. These BK equations are integro-differential equations, they can numerically straightforward to solve on a lattice. The impact parameter dependence of scattering amplitude, $N(\mathbf{r}, Y)$, is neglected, therefore the amplitude does not depend on angle, $N(\mathbf{r}, Y) = N(|\mathbf{r}|, Y)$. We can solve $N(r, Y)$ at discrete values of transverse dipole size. The BK equation can be viewed as a set of differential equations which can be solved by using GNU Scientific Library (GSL). The GSL computes the numerical integrals using adaptive integration routines, interpolates data points using the cubic spline interpolation,  and solves the differential equations based on Runge-Kutta method.

The initial condition is needed to solve the BK equation. In this study, we take the initial condition given by the McLerran-Venugopalan model (MV)\cite{MV}:
\be
N(r, Y=0) = 1 - \exp\left[-\frac{r^2Q_s^2(Y)}{4} \ln\left(\frac{1}{r^2\Lambda^2_{QCD}} +e\right)\right].
\ee
Note that as we are interested in the rapidity dependence of the amplitude and the DLA impact on the amplitude in the saturation region, the initial condition is not resumed when solving the NLL BK equation in current work.

For the qualitative properties of BK equation studied in this work, the precise value of the running coupling constant is not important. To specific, we evaluate the running coupling in terms of a popular one-loop expression
\be
\alpha_s(r^2) = \frac{4\pi}{\beta\ln\left(\frac{4C^2}{r^2\Lambda^2_{QCD}}\right)},
\ee
with $\beta = \frac{11}{3}N_c - \frac{2}{3}N_f$. The value of the fudge factor can be obtained by fitting to the HERA data, its value is taken from Ref.\cite{JL}. We freeze the coupling to a fixed value $\alpha(r_{fr}) = 0.75$ for larger dipole size, $r > r_{fr}$, to regularize the infrared behavior.

\begin{figure}[h!]
\setlength{\unitlength}{1.5cm}
\begin{center}
\epsfig{file=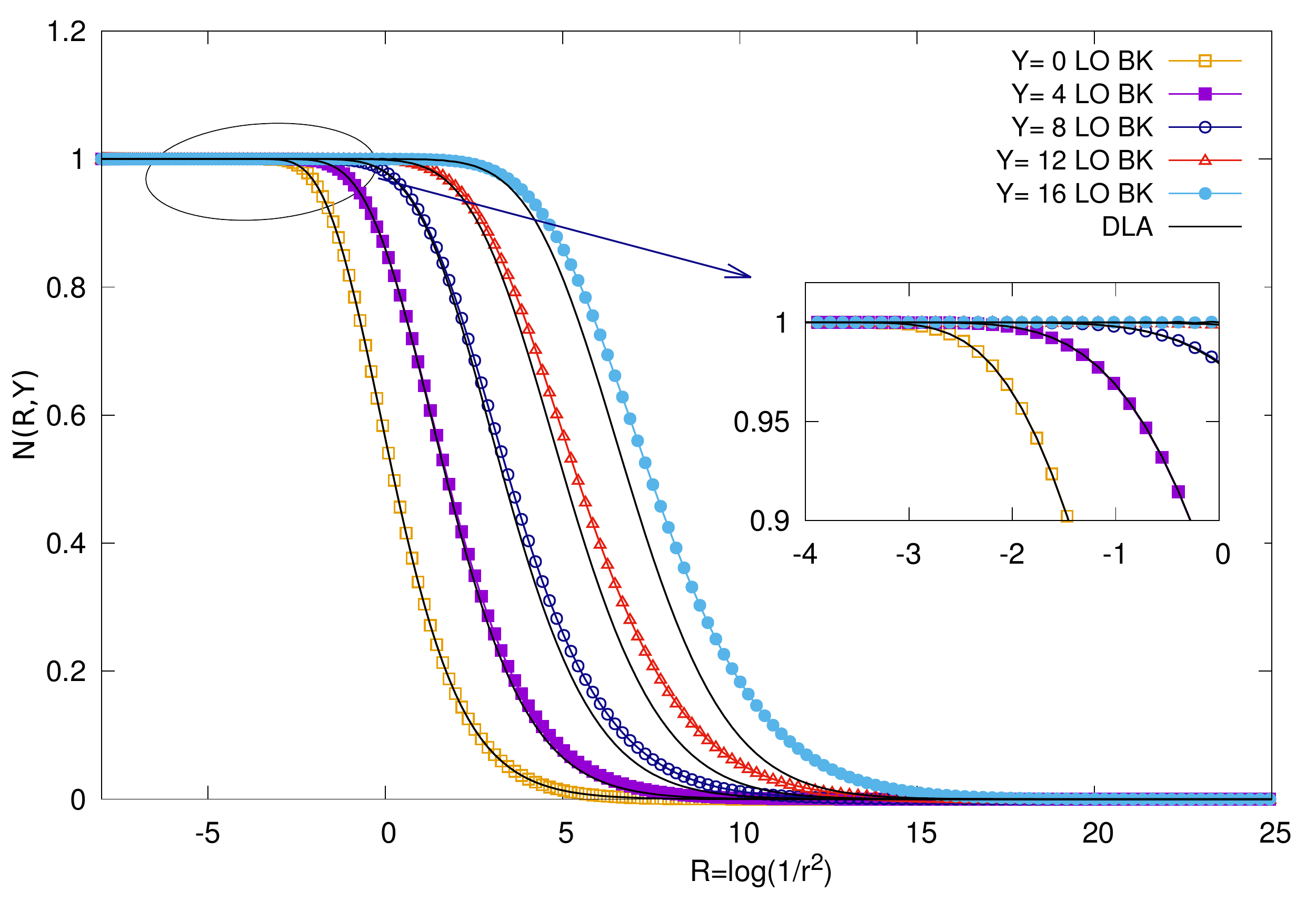, width=8cm,height=7cm}
\epsfig{file=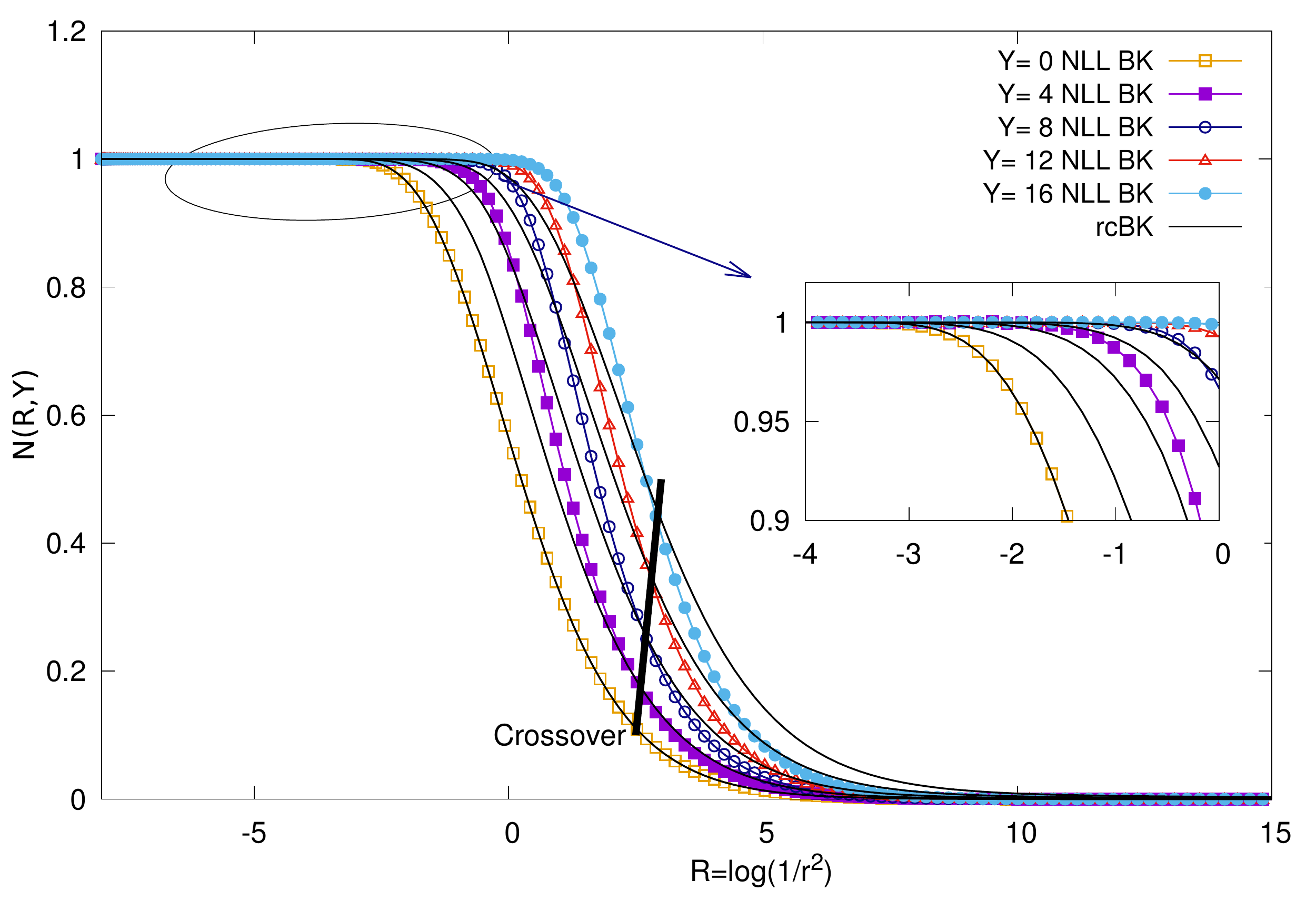, width=8cm,height=7cm}
\end{center}
\caption{Numerical solutions to the LO, DLA, rc and full NLL BK equations for 4 different rapidities. The inner diagrams are the zooming in scattering amplitude in the saturation region.}
\label{fignlo}
\end{figure}

The left hand panel of Fig.\ref{fignlo} shows the evolution of the dipole scattering amplitude with fixed coupling for 5 different rapidities. The evolution is faster for fixed coupling than the DLA BK equation in the non-saturation region, especially for large rapidities. While it seems that the DLA does not contribute to the evolution in the saturation region, as one can see the numerical LO and DLA results overlap each other from the inner zooming in diagram. These numerical outcomes are in agreement with the analytic results obtained in previous sections.

To demonstrate the change of rapidity dependence of scattering amplitude from running coupling to full NLL case, we numerically solve the NLL BK equation which includes corrections from full NLO and double logarithmic resummation. The right hand panel of Fig.\ref{fignlo} shows the solutions of rc and full NLL BK equations. The evolution is faster for the running coupling than the NLL in the non-saturation region where the dipole size is small and the scattering is weak. This outcome implies that the evolution is further slowed down on top of running coupling by the corrections from gluon loops and double logarithmic resummation in the non-saturation region. However, in the saturation region where the DLA can be neglected, we get opposite result, the evolution is slower for the running coupling than the NLL, one can see that the value of the amplitude in NLL larger than the one in running coupling case for the same rapidity from the inner diagram on the right hand panel of Fig.\ref{fignlo}. This result can be traced back to the gluon loop contributions in the full NLO case, see the second line in (\ref{fnlokernel}). The gluon loop corrections take the rapidity dependence of the exponent in $S$-matrix from linear to rapidity raised to the power of 3/2 (see Eqs.(\ref{solution_running}) and (\ref{sol_nll})), which indicate that the evolution in running coupling case is relative slow compared to the NLL one, but the evolutions for both of them are slower than the LO BK. There is a crossover between the saturation and the non-saturation regions, where the dipole size is around $\sim10^{-1}$. Beyond this crossover region, the evolution is dominated by the NLL corrections which slow down the evolution.

%-----------------------------------------------------------------------------

\begin{acknowledgments}
WX thanks the Department of Physics of The University of Colorado Boulder for the hospitality
during the early stages of this work. This work is supported by National Natural Science Foundation of China under Grant
No.11305040, the Education Department of Guizhou Province Talent Fund under Grant
No.[2015]5508, the Science and Technology Department of Guizhou Province Fund Nos.[2015]2114.
\end{acknowledgments}

%-----------------------------------------------------------------------------
%

%-----------------------------------------------------------------------------

\end{document}